\documentclass[aps,prl,twocolumn]{revtex4}
\usepackage{epsf}

\newcommand{\ybco}{YBa$_2$Cu$_3$O$_{7-\delta}$}

\newcommand{\bscco}{Bi$_2$Sr$_2$CaCu$_2$O$_{8+\delta}$}

\begin{document}

\title {
The pseudogap Fermi-Bose Kondo model
}

\author{Matthias Vojta and Marijana Kir\'{c}an}
\affiliation{\mbox{Theoretische Physik III, Elektronische Korrelationen und
Magnetismus, 
Universit\"at Augsburg, 86135 Augsburg, Germany}}
\date{October 12, 2002}

\begin{abstract}
We consider a magnetic impurity coupled to
both fermionic quasiparticles with a pseudogap density of states
and bosonic spin fluctuations.
Using renormalization group and large-$N$ calculations
we investigate the phase diagram of the resulting Fermi-Bose Kondo model.
We show that the Kondo temperature is
strongly reduced by low-energy spin fluctuations, and
make connections to experiments in cuprate superconductors.
Furthermore we derive an {\em exact} exponent for the
critical behavior of the conduction electron $T$ matrix,
and propose our findings to be relevant for certain scenarios
of local quantum criticality in heavy-fermion metals.
\end{abstract}

\maketitle


The interplay of quasiparticles and collective low-energy
excitations is a central theme in the physics of strongly
correlated materials.
Impurities have proven to be a powerful probe for investigating
the bulk behavior of such systems.
A paradigmatic model describing the interaction of impurity
degrees of freedom with both quasiparticles and collective spin fluctuations
is the so-called Fermi-Bose Kondo model \cite{bfk,bfknew},
consisting of a local impurity spin,
$\vec{S}$, coupled to spin-$\frac{1}{2}$ fermions and
spin-1 vector bosons.

The Fermi-Bose Kondo model has recently been analyzed for
the case of a metallic fermion density of states (DOS),
and a gapless bosonic spectrum representing magnetic order parameter fluctuations
at a bulk quantum critical point in $d$ dimensions \cite{bfk,bfknew}.
For $d<3$ the model shows a boundary quantum phase transition
between a Kondo-screened phase and a bosonic fluctuating phase with universal
local spin correlations.
Both the critical and the bosonic fluctuating fixed points are
characterized by singular thermodynamic properties,
connected with local non-Fermi liquid behavior.

The purpose of this Letter is to generalize this analysis to
a fermionic bath with a pseudogap DOS, $\rho_c(\omega)\propto |\omega|^r$,
and to discuss Kondo screening for the case of a finite spin gap,
$\Delta_s$, in the bosonic bath.
Using renormalization group (RG) methods, we shall establish the phase diagram
of the pseudogap Fermi-Bose Kondo model,
and furthermore derive an exact relation between the anomalous exponent
of the fermion $T$ matrix and the exponent of fermionic bath DOS, $r$, which
holds at any fixed point with a finite coupling between fermions and impurity
spin.
We shall apply our model to impurity moments in cuprate superconductors,
where the fermionic DOS obeys $r=1$, and show that Kondo screening is strongly
suppressed by collective spin fluctuations, in agreement with NMR
experiments.

Recently, the Fermi-Bose Kondo model has been
proposed \cite{edmft} to describe a ``local quantum phase transition'' in alloys like
$\rm CeCu_{6-x} Au_x$ \cite{schroder}.
This modelling is based on an extended dynamical mean-field
theory \cite{edmft}, where a lattice model is mapped onto
a self-consistent single-impurity model.
Based on our $T$ matrix analysis, we will argue that no anomalous power law
can occur in the fermion spectrum within such a scenario of local criticality.


{\it Model.}
The Fermi-Bose Kondo Hamiltonian for a spin-$\frac{1}{2}$ impurity can be written as
$H=H_{\rm f}+H_{\rm b}+H_{\rm imp}$, with
\begin{equation}
H_{\rm imp} = J_K {\vec S} \cdot {\vec s}_0 + \gamma_0 {\vec S} \cdot {\vec \phi}_0 \,.
\label{bfk}
\end{equation}
Here, $\vec S$ denotes the impurity spin operator, and
$H_{\rm f}$ describes a system of free fermions,
$H_{\rm f} = \sum_{{\bf k}\sigma} \epsilon_{\bf k} c^\dagger_{{\bf k}\sigma} c_{{\bf k}\sigma}$
in standard notation,
${\vec s}_0 = \sum_{\bf k k'} c^\dagger_{{\bf k}\sigma} {\vec \sigma}_{\sigma\sigma'}
c_{{\bf k '}\sigma'}$ is the conduction band spin operator at the impurity
site ${\bf r}_0 = 0$.
The fermionic density of states follows a power law at low energies,
$\rho_c(\omega)\!=\!\sum_{\bf k} \delta(\omega-\epsilon_{\bf k}) \propto N_0 |\omega|^r$,
which includes the cases of a metal ($r\!=\!0$) and a $d$-wave superconductor ($r\!=\!1$).
In the simplest case, the bosonic bath consists of free vector bosons,
$H_{\rm b} = \sum_{{\bf q}\alpha} \omega_{\bf q} b_{{\bf q}\alpha}^\dagger b_{{\bf q}\alpha}$,
describing collective spin fluctuations in the host material, with
a dispersion $\omega_{\bf q}^2 = m^2 + c^2 {\bf q}^2$.
At zero temperature, the mass $m$ coincides with the bulk spin gap $\Delta_s$.
A gap $\Delta_s>0$ describes a quantum paramagnet, $\Delta_s\!=\!0$ corresponds to a bulk quantum
critical point between the paramagnet and an antiferromagnetically ordered phase;
the momentum $\bf q$ is measured relative to the ordering wavevector
$\bf Q$.
The field ${\vec \phi}_0$ in Eq. (\ref{bfk}) represents the local orientation of the
antiferromagnetic order parameter, it is given by
$\phi_{0\alpha} = \sum_{\bf q} (b_{{\bf q}\alpha} + b_{-{\bf q}\alpha}^\dagger) /
\sqrt{\omega_{\bf q}/J}$, where $J$ is the bulk exchange constant.
It is useful to define a spectral density, $\rho_\phi(\omega)$, of the Bose field $\phi$.
At $T=0$ and for $\Delta_s=0$, it follows
$\rho_\phi(\omega) \propto K_0^2\, {\rm sgn}(\omega) |\omega|^{1-\epsilon}$,
$K_0^2=J/c^d$, in $d\!=\!3-\epsilon$ dimensions.
We note that in dimensions $d<3$,
boson self-interactions are relevant in the RG sense (in the absence of Landau damping),
and we will comment on their effect below.
Our analysis will be restricted to the SU(2)-invariant model;
the effect of spin anisotropy for $r\!=\!0$ has been discussed in
Ref.~\onlinecite{bfknew}.

In the absence of the bosonic bath, $\gamma\!=\!0$, and for $r\!=\!0$
the model (\ref{bfk}) reduces to the well-known Kondo model \cite{hewson},
which exhibits screening of the magnetic moment below a characteristic (Kondo)
temperature, $T_K$.
The pseudogap generalization, $r>0$, shows a zero-temperature
phase transition between a free local moment and a screened moment as the
Kondo coupling is varied \cite{withoff,bulla,GBI,OSPG}.
On the other hand, in the absence of the fermionic bath, $J_K\!=\!0$,
the Bose-Kondo model describes an impurity moment in an insulating quantum
antiferromagnet~\cite{science}.


{\it RG analysis.}
At the bulk quantum critical point we have $\Delta_s=0$, and both fermionic and
bosonic baths obey a power law DOS.
The impurity behavior can be analyzed using field-theoretic RG
methods, together with an expansion in $\epsilon$ and $r$,
similar to Refs.~\onlinecite{bfk,science,bfknew}.

We employ a pseudo-fermion representation of the impurity spin $\frac{1}{2}$,
${\vec S} = f^\dagger_\sigma {\vec\sigma}_{\sigma\sigma'} f_\sigma'$.
Following the standard procedure \cite{bgz}, we introduce renormalized
fields $f_\sigma = \sqrt{Z_f} f_{R\sigma}$, $\phi_\alpha = \sqrt{Z} \phi_{R\alpha}$,
and dimensionless couplings
$K_0 \gamma_0 = \mu^{\epsilon/2} [\widetilde{Z}_{\gamma}/(Z_f\sqrt{Z})] \gamma$,
$N_0 J_K = \mu^{-r} (Z_J / Z_f) j$,
where $\mu$ is a renormalization
energy scale. 
The derivation of the RG equations has been presented in Ref.~\onlinecite{bfknew}
for non-interacting $\vec\phi$ bosons ($Z=1$ in this case).
A similar procedure for the present case with a fermionic pseudogap DOS,
using the minimal subtraction scheme,
gives the following result for the RG beta functions:
\begin{eqnarray}
\beta(\gamma) &=& -\frac{\epsilon\gamma}{2} + \gamma^3 - \gamma^5 + \frac{j^2\gamma}{2} \,, \nonumber \\
\beta(j) &=& rj - j^2 + \frac{j^3}{2} + j\gamma^2 - j\gamma^4 \,.
\label{beta}
\end{eqnarray}
The full two-loop result for the case of interacting bosons will be given elsewhere.
The equations (\ref{beta}) reduce to the cases of the pseudogap Kondo model
for $\gamma=0$ \cite{withoff} and the metallic Bose-Fermi Kondo model for $r=0$ \cite{bfknew}.
\begin{figure}[!t]
\epsfxsize=2.4in
\centerline{\epsffile{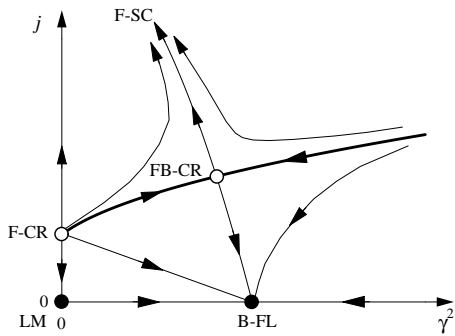}}
\caption{
Schematic RG flow diagram
for the pseudogap Fermi-Bose Kondo model
for small $\epsilon$ and $r$, and $\Delta_s\!=\!0$.
The axes denotes the couplings to the bosonic ($\gamma$) and
fermionic ($j$) baths.
The model has three (meta)stable fixed points (solid dots): the
local moment (LM), bosonic fluctuating (B-FL), and
fermionic strong-coupling (F-SC) fixed points;
the open dots are the critical fixed points F-CR and FB-CR;
for details see text.
The thick line corresponds to a line of continuous boundary phase transitions.
In the metallic case, $r=0$, the LM and F-CR fixed points
merge~\protect\cite{bfk,bfknew}.
}
\vspace*{-10pt}
\label{fig:flow}
\end{figure}
Besides the decoupled (local moment -- LM) fixed point, $\gamma^\ast\!=\!j^\ast\!=\!0$,
the above flow equations have three non-trivial fixed points which are
perturbatively accessible:

\noindent
B-FL: ${\gamma^\ast}^2 = \frac{\epsilon}{2} + \frac{\epsilon^2}{4}$, $j^\ast=0$
corresponds to a stable fixed point where the fermionic Kondo coupling is irrelevant,
and the coupling between impurity and bosonic bath leads to universal local-moment
fluctuations \cite{science};

\noindent
F-CR: $j^\ast = r + \frac{r^2}{2}$, $\gamma^\ast=0$ is an unstable fixed point
controlling the transition between the decoupled (free) moment and the Kondo-screened
moment phases; and

\noindent
FB-CR: ${\gamma^\ast}^2 = \frac{\epsilon}{2} + \frac{\epsilon^2}{8} - \frac{\epsilon r}{2} -
\frac{r^2}{2}$, $j^\ast = r + \frac{\epsilon}{2}$ -- this is an unstable fixed point
between the bosonic fluctuating and the Kondo-screened phases.

For large enough $j$, $\gamma$ is irrelevant, and the model reduces to the fermionic
pseudogap Kondo model which shows strong-coupling phases with screening.
Numerical work \cite{GBI}
has shown that, depending on the value of the exponent $r$ and
the presence or absence of particle-hole symmetry, two different strong-coupling fixed
points exist.
As these are not accessible within the present perturbation expansion, we will
not distinguish between them, and simply denote the fermionic strong-coupling fixed
point, reached for large $j$ and small $\gamma$, with F-SC.
The flow diagram resulting from this discussion is shown in Fig.~\ref{fig:flow}.
In the presence of bosonic interactions the RG flow is similar, and
the location of the fixed points is modified only at order $\epsilon^2$.
We expect the general structure of the phase diagram, Fig.~\ref{fig:flow},
to be valid for all $\epsilon$ and $r$ values between 0 and 1 -- this is e.g.
the result of large-$N$ calculations (see below); for $r>\frac{1}{2}$
particle-hole asymmetry is needed to reach the F-SC fixed point.
Finally, a finite host spin gap, $\Delta_s>0$, will cut-off the RG flow
of $\gamma$ at the scale $\Delta_s$, but there is still a transition
as function of $J_K$ controlled by the F-CR fixed point.

\begin{figure}
\epsfxsize=3.2in
\centerline{\epsffile{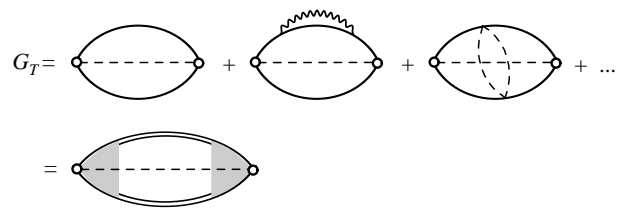}}
\caption{
Feynman diagrams for the Green's function $G_T$ entering the conduction
electron $T$ matrix. Full/dashed/wiggly lines denote $f$/$c$/$\phi$ propagators.
The double line is the full $f$ propagator, the shaded area the full vertex,
and the open dots are the external sources.
}
\label{fig:dgr}
\vspace*{-10pt}
\end{figure}

{\it $T$ matrix.}
An important quantity 
is the conduction electron $T$ matrix, describing the scattering of
the $c$ electrons off the impurity.
It is useful to define a propagator, $G_T$, of the composite
operator $T_\sigma = f_\sigma^\dagger f_{\sigma'} c_{\sigma'}$,
such that the $T$ matrix is given by $T(\omega) = J_K^2 G_T(\omega)$.
The diagrams for $G_T$ in perturbation theory are shown in
Fig.~\ref{fig:dgr}.
At the RG fixed points, we expect a power law behavior of the
$T$ matrix spectral density, with an anomalous
exponent $\eta_T$.
(Note that at tree level $G_T \propto\omega^r$ in the present pseudogap
problem.)
Within the standard scheme \cite{bgz}
we can obtain the $G_T$ renormalization factor $Z_T$ in an expansion in
$\gamma$ and $j$ by evaluating the lowest-order diagrams and demanding the
resulting expression to be free of poles at energy $\mu$.
From $Z_T$ we obtain a perturbative expression for $\eta_T$ according to
$\eta_T = \partial(\ln Z_T)/ \partial(\ln\mu)$, evaluated at $j=j^\ast$ and
$\gamma=\gamma^\ast$.

However, it is possible to obtain an {\em exact} result for $\eta_T$, valid
to all orders in perturbation theory.
The argument parallels the one for anomalous exponent of the local
susceptibility in Refs.~\cite{bfknew,science}:
Inspecting the diagrams in the upper panel of Fig.~\ref{fig:dgr} shows that all
are obtained from the first diagram with {\em full} $j$ vertices and
{\em full} $f$ electron propagators, i.e., there are no additional
processes involving, e.g., the conduction electron line.
This gives the relation $Z_T^{-1} = Z_J^2 / Z_f^2$~\cite{znote}.
Using the definition of $Z_J$ and taking the derivative w.r.t. $(\ln\mu)$
at fixed bare $J_K$, we obtain $0=-r + \beta(j)/j - \eta_T/2$.
Thus, at any RG fixed point with finite, non-zero $j^\ast$, we
find the exact result
\begin{equation}
\eta_T=-2r ~~\Rightarrow~~
T(\omega) \propto \frac{1}{\omega^{-r-\eta_T}} = \omega^{-r}.
\label{exact}
\end{equation}
This result applies to both critical fixed points
(F-CR and FB-CR), and relies only
on the fact that the $c$ electrons do not have a self-interaction
being relevant in the RG sense, i.e.,
it holds for interacting $\vec\phi$ bosons as well.

Remarkably, the critical behavior $T(\omega)\propto \omega^{-r}$ has been
numerically found in studies of the pseudogap Kondo model, i.e.,
at the F-CR fixed point \cite{bulla,MVRB}.
It is also obtained within a dynamic large-$N$ analysis of the same
problem \cite{OSPG}.


{\it Application to cuprate superconductors.}
In superconducting cuprates it is reasonable to separate the low-energy degrees
of freedom into Bogoliubov quasiparticles around the
nodes of the $d$-wave superconducting order parameter, and
antiferromagnetic spin fluctuations around wavevector ${\bf Q}=(\pi,\pi)$,
as manifest in the resonance peak in neutron scattering \cite{respeak};
momentum conservation prohibits a low-energy coupling between fermionic and
bosonic degrees of freedom.
A phenomenological description of cuprates in terms of a BCS superconductor
plus low-energy collective spin excitations~\cite{demler} has been successful in
describing a number of recent neutron scattering and STM experiments.
Therefore, impurity moments in cuprate $d$-wave superconductors can be expected
to be well described by the pseudogap Fermi-Bose Kondo model, with
fermionic DOS exponent $r\!=\!1$.

Experimentally, signatures of Kondo physics have been observed in
cuprates doped e.g. with Zn replacing Cu \cite{alloul,bobroff} --
here the non-magnetic Zn induces a quasi-free magnetic moment in its
vicinity,
which can be related to the properties of the parent Mott
insulator~\cite{fink}.
In particular, NMR experiments indicate Kondo screening of Zn-induced moments in
optimally doped \ybco\ below the superconducting $T_c$, but show that screening is
suppressed in underdoped samples \cite{bobroff}.
So far, theoretical analysis has been restricted to either
a moment interacting with bulk spin fluctuations \cite{science} or a
moment interacting with Bogoliubov quasiparticles \cite{withoff,MVRB,tolya}.

\begin{figure}
\epsfxsize=2.7in
\centerline{\epsffile{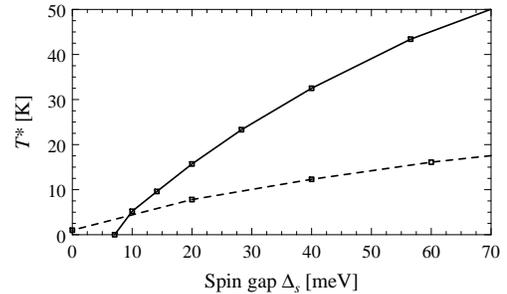}}
\caption{
Characteristic energy scale $T^\ast$ as function of the bosonic spin
gap $\Delta_s$, calculated for a fermionic host
corresponding to a cuprate superconductor \cite{MVRB},
using large-$N$ techniques.
Solid: $J_K\!=\!0.27$ eV, $U\!=\!-0.15$ eV, $\gamma_0=0.15$ eV;
Dash-dot: $J_K\!=\!0.38$ eV, $U\!=\!0$,
where $U$ is the additional impurity scattering potential.
In comparison with NMR experiments,
$T^\ast$ is equivalent to the Kondo temperature below which
the impurity moment is screened; the horizontal axis
can roughly be identified with doping (up to $\Delta_s \approx$ 50 meV) \cite{respeak}.
Note that the change in the fermionic band
upon doping has little influence on $T^\ast$;
an increase in the superconducting gap, $\Delta_0$, with underdoping
will further suppress $T^\ast$.
}
\label{fig:tk1}
\vspace*{-10pt}
\end{figure}

Here we propose that the increasing strength of
antiferromagnetic fluctuations is the most important contribution to the $T_K$
suppression in underdoped cuprates.
To this end, we calculate the Kondo temperature in the pseudogap
Fermi-Bose Kondo model for a realistic band structure in $H_{\rm f}$
and a spin fluctuation spectrum in $H_{\rm b}$ with varying spin
gap $\Delta_s$.
In modelling the experiment, the important ingredient is that
$\Delta_s$ decreases with underdoping, as observed in neutron
scattering \cite{respeak} --
from the RG analysis it is clear that decreasing $\Delta_s$ can drive the
impurity from the F-SC to the LM phase.
For $H_{\rm f}$ we use a standard $d$-wave BCS model with
a gap function $\Delta_{\bf k} = (\Delta_0/2) (\cos k_x - \cos k_y)$.

To obtain quantitative results, we employ a large-$N$ approach were
the spin symmetry is generalized from SU(2) to SU($N$).
The methodology is similar to earlier work~\cite{hewson,science,tolya,burdin},
details will be published elsewhere.
While this method has artifacts near the quantum-critical points
discussed above, and is not quantitatively accurate, we expect it to capture
the qualitative low-$T$ physics.
The $T=0$ results allow to extract a low-energy scale, namely the
position of the peak in the $T$ matrix, $k_B T^\ast$, which can be identified
with the Kondo temperature up to a numerical factor of order unity \cite{tolya,MVRB}.
Fig.~\ref{fig:tk1} shows sample results for this quantity as function of the
bulk spin gap, $\Delta_s$.

These results indicate that antiferromagnetic fluctuations with a small
spin gap $\Delta_s$ are effective in suppressing Kondo screening.
We note that an increasing superconducting gap, $\Delta_0$, upon underdoping
can have a similar effect in the pseudogap Kondo model~\cite{MVRB}.
However, this mechanism is less likely to be universal for all cuprates,
as a strong doping dependence of $\Delta_0$ has been observed
in \bscco\ but not in \ybco.
In contrast, spin fluctuations grow upon underdoping in all cuprates.
Therefore, we propose the suppression of Kondo screening as observed in
NMR \cite{bobroff} to be well explained within the
pseudogap Fermi-Bose Kondo model with a doping-dependent host
spin gap.


{\it Application to heavy-fermion metals.}
Motivated by neutron scattering experiments on the
heavy-fermion compound $\rm CeCu_{6-x} Au_x$ \cite{schroder},
which indicate momentum-independent critical dynamics at
an antiferromagnetic ordering transition,
Si {\em et al.} proposed a self-consistent version of the Fermi-Bose
Kondo model as a candidate to describe such critical
behavior within an extended dynamical mean-field
approach (EDMFT) \cite{edmft}.
In this scenario, the critical point of the lattice model is
mapped onto the critical point (BF-CR) of the
impurity model.

The self-consistency within EDMFT has to be achieved for both
fermionic and bosonic baths.
So far, most discussions assumed $r\!=\!0$ and focussed primarily on the bosonic
part \cite{edmft,bfknew}.
Here we want to draw attention to the fermionic part as
well.
Our analysis above shows that a fermionic bath with $|\omega|^r$
spectrum leads to a impurity $T$ matrix with power-law divergence $|\omega|^{-r}$ (\ref{exact})
at the critical point.
Now the question is whether DMFT self-consistency in the fermionic
bath of a Kondo or Anderson lattice model can be achieved with such a
power law.
Consider the DMFT self-consistency relation for an Anderson
lattice model, where the $T$ matrix corresponds to $V^2 G_{\rm imp}$ \cite{hewson}:
\begin{equation}\label{sceq}
G^{\rm loc}(z)\begin{array}[t]{@{\;=\;}l}\displaystyle\int
d\epsilon\frac{\rho_0^c(\epsilon)}{z-\varepsilon_f-\Sigma^f(z)-\frac{V^2}{z-\epsilon}}\\[5mm]
[z-\varepsilon_f-{\Delta}(z)-\Sigma^f(z)]^{-1} = G_{\rm imp}(z)
\end{array}
\end{equation}
in standard notation,
with $\rho_0^c(\epsilon)$ the DOS of the bare conduction
band of the lattice model, $\Sigma^f$ the interacting impurity self-energy,
and ${\Delta}(z)$ the effective hybridization function for the impurity
model.
Now, $G_{\rm imp} \propto \omega^{-r}$ implies
${\rm Im}[\Delta(\omega)+\Sigma^f(\omega)] \propto -|\omega|^{r}$.
As $\Sigma^f$ is identified with the lattice self-energy within DMFT,
we require ${\rm Im}\Sigma^f < 0$, and both ${\rm Im}\Delta(\omega)$
and ${\rm Im}\Sigma^f(\omega)$ have to vanish as $-|\omega|^{r}$. 
Using the ansatz $\Sigma^f(z) = -\varepsilon_f + A z^r$ and
expanding $\rho_0^c(\epsilon)$ around $\epsilon\!=\!0$,
the integral in the first line of (\ref{sceq}) can be performed in the
limit of small $z$, and, surprisingly, the coefficient of the
leading $z^{-r}$ term in $G^{\rm loc}(z)$ vanishes.

Therefore, critical-point self-consistency in the fermionic sector of EDMFT
for an Kondo or Anderson model {\em cannot} be achieved with an exponent $r>0$.
If a local critical point within EDMFT exists, it is thus characterized by
$r\!=\!0$, and in this case logarithmic corrections in the $T$ matrix are
expected, leading to a logarithmic temperature dependence of the
resistivity.
In addition, we note that any such critical point will be characterized
by an extensive ground state entropy (this holds for arbitrary $r$),
which should have strong signatures in specific heat experiments.


{\it Conclusions.}
We have analyzed a Fermi-Bose Kondo model, describing an impurity spin
coupled to both fermionic quasiparticles with a pseudogap DOS and collective
mode bosons, using renormalization group and large-$N$ techniques.
We have argued our findings to be relevant for cuprate superconductors
as well as for scenarios of local criticality in heavy-fermion
metals.



We thank R. Bulla, A. Polkovnikov, Th. Pruschke, and S. Sachdev for
collaboration on related work
and H. Alloul, S. Burdin, A. Rosch, and Q.~Si
for discussions.
This research was supported by the DFG SFB 484.



\vspace*{-10pt}

\begin{thebibliography}{}

\vspace*{-10pt}

\bibitem{bfk}
J.~L.~Smith and Q.~Si, 
Europhys. Lett. {\bf 45}, 228 (1999);
A.~M.~Sengupta, Phys. Rev. B {\bf 61}, 4041 (2000).

\bibitem{bfknew}
L. Zhu and Q. Si, Phys. Rev. B {\bf 66}, 024426 (2002);
G. Zarand and E. Demler, Phys. Rev. B {\bf 66}, 024427 (2002).

\bibitem{edmft} Q.~Si, S.~Rabello, K.~Ingersent, and J.~L.~Smith, Nature
{\bf 413}, 804 (2001) and cond-mat/0202414.

\bibitem{schroder}
A. Schr\"oder {\em et al.},
Nature {\bf  407},   351 (2000).

\bibitem{hewson} A.~C.~Hewson,
{\em The Kondo Problem to Heavy Fermions}, Cambridge
University Press, Cambridge (1997).

\bibitem{withoff}
D.~Withoff and E.~Fradkin, Phys. Rev. Lett. {\bf 64}, 1835 (1990).
%
C.~R.~Cassanello and E.~Fradkin,
Phys. Rev. B {\bf 53}, 15079 (1996) and {\bf 56}, 11246 (1997).

\bibitem{bulla} R.~Bulla, T.~Pruschke,
and A.~C.~Hewson, J. Phys.: Condens. Matter {\bf 9}, 10463 (1997),
R.~Bulla, M.~T.~Glossop, D.~E.~Logan, and T.~Pruschke,
{\em ibid} {\bf 12}, 4899 (2000).

\bibitem{GBI}
C.~Gonzalez-Buxton and K.~Ingersent, \prb {\bf 57}, 14254 (1998);
K.~Ingersent and Q.~Si, Phys. Rev. Lett. {\bf 89}, 076403 (2002).

\bibitem{OSPG} M. Vojta, \prl {\bf 87}, 097202 (2001).


\bibitem{science} S.~Sachdev, C.~Buragohain and M.~Vojta, Science
{\bf 286}, 2479 (1999); M.~Vojta, C.~Buragohain and S.~Sachdev,
Phys. Rev. B {\bf 61}, 15152 (2000).

\bibitem{bgz}
E. Brezin, J.C. Le Guillou, and J. Zinn-Justin, in {\em Phase
transitions and critical phenomena}, eds.
C. Domb and M. S. Green, Page Bros., Norwich (1996), Vol. 6.

\bibitem{znote}
A relation between $Z$ factors can only be formulated for a
true propagator like $G_T$, not for a self energy.

\bibitem{MVRB} M. Vojta and R. Bulla, \prb {\bf 65}, 014511 (2002).



\bibitem{respeak}
H. F. Fong {\em et al.}, \prb {\bf 61}, 14773 (2000).

\bibitem{demler}
E.~Demler, S.~Sachdev, and Y.~Zhang, \prl {\bf 87}, 067202 (2001);
A.~Polkovnikov, M.~Vojta, and S.~Sachdev, Phys. Rev. B {\bf 65}, 220509 (2002).

\bibitem{alloul}
H. Alloul {\em et al.}
\prl {\bf 67}, 3140 (1991).

\bibitem{bobroff} J.~Bobroff {\em et al.}, Phys. Rev. Lett. {\bf
83}, 4381 (1999) and {\bf 86}, 4116 (2001).







\bibitem{fink}
A.~M.~Finkelstein {\em et al.},
Physica C {\bf 168}, 370 (1990);
S. Sachdev and M. Vojta,
cond-mat/0009202.

\bibitem{tolya} A.~Polkovnikov, S.~Sachdev, and M.~Vojta, \prl {\bf 86}, 296 (2001).

\bibitem{burdin}
S. Burdin {\em et al.}, \prb {\bf 67}, 121104 (2003). 


\end{thebibliography}
\end{document}